\documentclass[12pt]{article}
\usepackage{amssymb,amsmath,graphicx}
\begin{document}
\title{\textbf{Gribov copies and topological charge}} 
\author{B.~Holdom%
\thanks{bob.holdom@utoronto.ca}\\
\emph{\small Department of Physics, University of Toronto}\\[-1ex]
\emph{\small Toronto ON Canada M5S1A7}}
\date{}
\maketitle
\begin{abstract}
The existence of Gribov copies is a central feature of the field configuration space of confining gauge theories. In particular a transition between two Gribov copies with relative winding number implies a space-time configuration with topological charge. We explicitly demonstrate the proliferation of Gribov copies with relative winding number, where our focus is on localized (finite norm) configurations in Coulomb gauge. We then discuss the likelihood that some pairs of such copies are connected by Minkowski space solutions. We also comment on the relative importance of instantons and the connection to confinement.
\end{abstract}

\section{Introduction}
Instantons and sphalerons are examples in gauge theories of transitions that carry topological charge. Instantons can be interpreted as quantum tunnelling events between degenerate vacua \cite{a9}, while sphalerons are configurations at the top of a barrier separating degenerate vacua \cite{a10}. Instantons are vacuum fluctuations while sphalerons are relevant to the high temperature fluctuations over the barrier. In both cases the gauge field configurations characterizing the degenerate vacua are pure-gauge. The associated gauge transformation that relates these pure-gauge fields carries a winding number, and the result of the transition is an incontractible loop in configuration space.

But gauge theories are also characterized by gauge equivalent configurations that are not pure-gauge, even after gauge fixing. These are known as Gribov copies \cite{a1}, and their appearance in covariant gauges such as Landau gauge, similar gauges such as Coulomb gauge, and temporal gauge (once Gauss's law is enforced \cite{a12}) is a reflection of the nontrivial topology of configuration space. In particular gauge transformations relating different Gribov copies can also carry winding number. If the fields on two different time slices are related in this way, then a spacetime configuration that interpolates between these Gribov copies will carry topological charge. The result is again an incontractible loop in configuration space. It is of interest to know whether these transitions could play a role in the dynamics of topological fluctuations in gauge theories.

We shall work in Coulomb gauge, $\partial_i A_i^a=0$. In this gauge the field configurations are stationary points of the norm functional $||A||\equiv\int d^3x A^a_iA^a_i$ under gauge transformations. Gribov copies correspond to more than one stationary point of the norm on a gauge orbit. We shall elevate the significance of the norm further by requiring that gauge field configurations have finite norm.

Finite norm has several consequences for our study.\footnote{The requirement of finite norm is more than is strictly necessary. Fields with slower fall-off at large $r$ can still give (1) and (2) and to a lesser extent (3).} 1) It allows the points at spatial infinity to be identified, giving the time slice the topology of $S^3$. This then guarantees the existence of Gribov copies, since a compact manifold forms the basis\footnote{It excludes axial gauge for instance.} of the proof by Singer \cite{a11} that there exists no gauge-fixing scheme that is free of Gribov copies. 2) The Gribov copies occur discretely and in finite numbers, at least when spherical symmetry is also imposed. Then Gribov copies can be explicitly counted and their relative winding numbers for instance can be directly obtained \cite{a2}.  3) The field configurations correspond to localized fluctuations, in the sense that spatially separated fluctuations may have negligible mutual influence. Thus our discussion of a single isolated topological fluctuation may still be relevant even when it is occurring in the midst of a whole sea of such fluctuations. And this sea, the vacuum, can have Lorentz invariance, even though our description of a single fluctuation does not.

Our focus then is on smooth configurations that are localized in space and where the topological charge is also localized in time (finite time duration). We shall consider the question of whether configurations of this type can also be classical solutions in real time. Minkowski space solutions carrying topological charge that have been considered before are quite different. In \cite{a5} solutions take the form of spherical shells of energy which come in from infinity and then go back out, leaving behind a change in topological charge. So these solutions are localized in neither space or time. In \cite{a14} the configurations carrying topological charge are physically static, with a periodic time dependence carried by a time dependent gauge transformation. These configurations have infinite norm on spatial slices, which occurs when a gauge orbit connects configurations of different winding number. As will become more clear below, only when the connecting path involves gauge inequivalent configurations can the configurations have finite norm.

The configuration space of finite norm gauge fields contains within it a ``fundamental modular region'' (FMR), defined such that its interior is free of Gribov copies. The FMR is the set of configurations that attain the global minimum of the norm on each gauge orbit. Nevertheless there are configurations that are gauge equivalent to each other located on the surface of the FMR, with the added property that these identified points have degenerate norm \cite{a3}. These identified points can have relative winding numbers, and thus a path in the FMR that begins and ends on two such points is an incontractible loop. The paths mentioned earlier connecting two Gribov copies can be gauge transformed into such  loops in the FMR. This illustrates how the nontrivial topology is an intrinsic property of the configuration space.

To clarify further, we have referred to a path in configuration space that carries topological charge as an incontractible loop. Somewhere along the loop is a discontinuous jump between two gauge equivalent configurations with relative winding number. We have placed these two configurations on different time slices so that the interpolating spacetime configuration can be smooth. A gauge transformation moves the location of an incontractible loop in configuration space, and a gauge transformation with winding number acting on some section of the loop can be used move the location of the discontinuity to a different, inequivalent point on the loop.

As an example we can take a normal instanton configuration and transform it to Coulomb gauge and finite norm. In this case there is no pure-gauge configuration except $A_\mu^a=0$, as described below, and thus the vacuum to vacuum transition must begin and end with this. But this means that there must be a discontinuity somewhere during the interpolating spacetime configuration, where a jump from one Gribov copy to another occurs. This agrees with the results of \cite{a13}. If the instanton is moved to lie entirely within the FMR then the discontinuous jump occurs between configurations of degenerate norm. But of course it is more convenient to consider instantons in other gauges where discontinuities do not occur.

We should also mention that the difficulties of explicitly specifying the FMR has meant that a larger region is usually considered instead. This is the ``first Gribov region'' \cite{a1} and it is obtained by requiring that each allowed configuration is only a local minimum of the norm. This region also contains Gribov copies \cite{a4}. We do not consider this region or the FMR further, since the information contained in the topological intricacies of the FMR will be reflected in the properties of Gribov copies in the encompassing and more easily defined space of finite norm gauge fields. And as described in \cite{a1}, Gribov copies can be accounted for in a path integral that is defined on this larger space of fields.

\section{Gribov copies and winding number}
We take the gauge group to be $SU(2)$ with gauge field $A_\mu(x)=A^a_\mu(x)\sigma^a/2$, defined so that the action is $S=\frac{1}{4g^2}\int d^4xF_{\mu\nu}^aF^{a\mu\nu}$. We begin with general time independent spherically symmetric gauge configurations of the form
\begin{eqnarray}
A^a_0&=&\frac{f_0(r)}{r}\frac{x_a}{r},\\
A^a_i&=&\frac{f_1(r)}{r}\frac{\varepsilon_{iab}x^b}{r}+\frac{f_2(r)}{r}\left(\delta_{ai}-\frac{x_a x_i}{r^2}\right)+\frac{f_3(r)}{r}\frac{x_a x_i}{r^2}
.\label{e2}\end{eqnarray}
The Coulomb gauge fixing condition $\partial_i A_i^a=0$ gives
\begin{equation}
(r f_3(r))'=2f_2(r).
\label{e1}\end{equation}
Finite norm implies
\begin{equation}
||A||=4\pi\int_0^\infty dr \left[2f_1^2+2f_2^2+f_3^2\right]<\infty
.\label{e6}\end{equation}

The following gauge transformations, a $U(1)$ subgroup of $SU(2)$, preserve the spherical symmetry,
\begin{equation}
U(x)=\cos(\alpha(r)/2)\mathbf{1}+i\sin(\alpha(r)/2)\frac{x^a}{r}\sigma^a
.\label{e3}\end{equation}
For this to be consistent with the identification of points at spatial infinity, we must require $U(x)\rightarrow\pm\mathbf{1}$ as $r\rightarrow\infty$, or in other words $\alpha(\infty)=2n\pi$ for integer $n$. To be a smooth transformation $U(x)\rightarrow\pm\mathbf{1}$ as $r\rightarrow0$ as well. Without lack of generality we can set $\alpha(0)=0$, in which case $n$ labels the winding number of the gauge transformation.

The transformation $A\rightarrow A^U$ reads
\begin{eqnarray}
f_0(r)&\rightarrow& f_0(r),\\
1+f_1(r)&\rightarrow& \cos(\alpha(r))(1+f_1(r))+\sin(\alpha(r))f_2(r),\label{e7}\\
f_2(r)&\rightarrow& \cos(\alpha(r))f_2(r)-\sin(\alpha(r))(1+f_1(r)),\\
f_3(r)&\rightarrow& f_3(r)-r\alpha'(r).
\end{eqnarray}
For $A^U$ to be a Gribov copy it must satisfy $\partial_i(A^U)_i^a=0$, and this leads to the Gribov pendulum equation,
\begin{equation}
r^2\alpha''(r)+2r\alpha'(r)-2(1+f_1(r))\sin(\alpha(r))+2f_2(r)(\cos(\alpha(r))-1)=0
,\label{e4}\end{equation}
where $f_3$ has been eliminated using (\ref{e1}). Solutions of this equation are parameterized by $\alpha'(0)$, and we must seek those solutions with $\alpha(0)=0$ and $\alpha(\infty)=2n\pi$.

We consider the Chern-Simons charge and related quantities,
\begin{equation}
Q=\int d^3x J_0,\quad\quad\partial_\mu J^\mu=\frac{1}{32\pi^2}\varepsilon^{\mu\nu\alpha\beta}F_{\mu\nu}^a F_{\alpha\beta},
\end{equation}
\begin{equation}
J^\mu=\frac{1}{16\pi^2}\varepsilon^{\mu\alpha\beta\gamma}[A^a_\alpha \partial_\beta A^a_\gamma+\frac{1}{3}\varepsilon_{abc}A^a_\alpha A^b_\beta A^c_\gamma]
\label{e5}.\end{equation}
The value of $Q$ for a configuration at some time $t$ typically has no special significance and it need not be integer valued. But a gauge transformation of winding number $n$ changes it by $Q\rightarrow Q+n$. Now suppose that two finite norm gauge configurations at time $t_i$ and $t_f$ are gauge equivalent via a gauge transformation with winding number $n$. Then the spacetime configuration which interpolates between the fields on the two time slices will carry an integer topological charge $q=Q(t_f)-Q(t_i)=n$, where
\begin{equation}
q=\frac{1}{32\pi^2}\int_{t_i}^{t_f}dt\int d^3x\;\varepsilon^{\mu\nu\alpha\beta}F_{\mu\nu}^a F_{\alpha\beta}
.\end{equation}

In the normal description of an instanton the configurations at $t=-\infty$ and $t=\infty$ are pure gauge, each with an integer value of $Q$. The instanton tunnels between different vacuum sectors characterized by these integer values. But it is well known that this picture does not survive in Coulomb gauge \cite{a7,a13}. The only nontrivial pure-gauge ($f_i=0$) solutions of (\ref{e4}) have $\alpha(\infty)=\pm \pi$, and this corresponds to half-integer $Q=\pm\frac{1}{2}$. There is a continuum of such solutions existing for any $\alpha'(0)\neq0$, with sign($\alpha(\infty)$)$=$sign($\alpha'(0)$). But these solutions are not consistent with the identification of points at infinity, and they have infinite norm ($f_1(\infty)=-2$ from (\ref{e7})). Thus only the trivial vacuum configuration remains, $A_\mu^a=0$.

For fields of small amplitude ($f_i\approx0$) the Gribov pendulum equation continues to have no nontrivial solutions that maintain finite norm. Our gauge field definition is such that $A_\mu^a=g\tilde{A}_\mu^a$ where $\tilde{A}_\mu^a$ has canonical kinetic terms, and thus the $f_i$'s have implicit factors of the gauge coupling. In weakly coupled gauge theories the $f_i$'s are effectively small and Gribov copies are not expected to play a role. For fields of larger amplitude there can be solutions with $\alpha(\infty)=2n\pi$, as we shall discuss. Thus, as is well known, only in strongly interacting gauge theories can the normal vacuum fluctuations be expected to feel the presence of Gribov copies.

We first discuss finite norm Gribov copies for certain special configurations. When $f_1\neq0$ and $f_2=f_3=0$ it can be deduced from (\ref{e5}) that $Q=0$, and thus the Gribov copies found in this case have integer values of $Q$. The solutions to (\ref{e4}) in this case were investigated in \cite{a15,a16}. In \cite{a2} a method was developed for the explicit counting of such Gribov copies for various $Q$. The number of Gribov copies was found to be sensitive to the product of the gauge field amplitude and its typical wavelength, with the number growing quickly from zero when this product is above some critical value. The set of the possible winding numbers $n$ among a set of copies was also sensitive to the form and amplitude of $f_1$. The distribution of copies as a function of $\alpha'(0)$ was found to have properties of interest to confinement.

We may also obtain finite norm Gribov copies with half-integer $Q$. Consider smooth configurations with $f_1\neq0$ and $f_2=f_3=0$ where $f_1(r)$ interpolates between 0 and $-2$ as $r$ varies from 0 to $\infty$. These are similar to the infinite norm configurations mentioned above, but now we consider configurations that are not pure-gauge. One then looks for solutions of (\ref{e4}) with $\alpha(r)\rightarrow (n+\frac{1}{2})2\pi$ as $r\rightarrow\infty$. Such a gauge transformation would transform the configuration with $Q=0$ and infinite norm to one with $Q=n+\frac{1}{2}$ and finite norm ($f_1(\infty)=0$ from (\ref{e7})). We find that such solutions do exist and that there can be different solutions having different values of $n$, for a given $f_1$. These gauge configurations with half-integer $Q$ are reminiscent of sphaleron configurations, with the added restriction of Coulomb gauge.

These examples of Gribov copies having integer or half-integer $Q$ are sets of measure zero relative to the general case, where typically all the $f_i$ are nonvanishing and the $Q$ values are unconstrained. In general only the differences $\Delta Q$ between copies have integer values. The copies and their winding numbers are sensitive to the form and amplitude of both functions $f_1$ and $f_2$ (with $f_3$ determined by (\ref{e1})). We can explore the solutions to the Gribov pendulum equation in the general case as a function of $\alpha'(0)$, and thus find the values that give $\alpha(\infty)=2\pi n$. Fig.~(1) shows the existence of such finite norm Gribov copies for one example choice of the $f_i$. We see again that there can be multiple copies for each of the possible values of $n$, and that both the number of copies and the range of $n$ values is finite. The numbers increase rapidly with the amplitude of the $f_i$'s.
\begin{figure}[t]
\centering\includegraphics[scale=0.6]{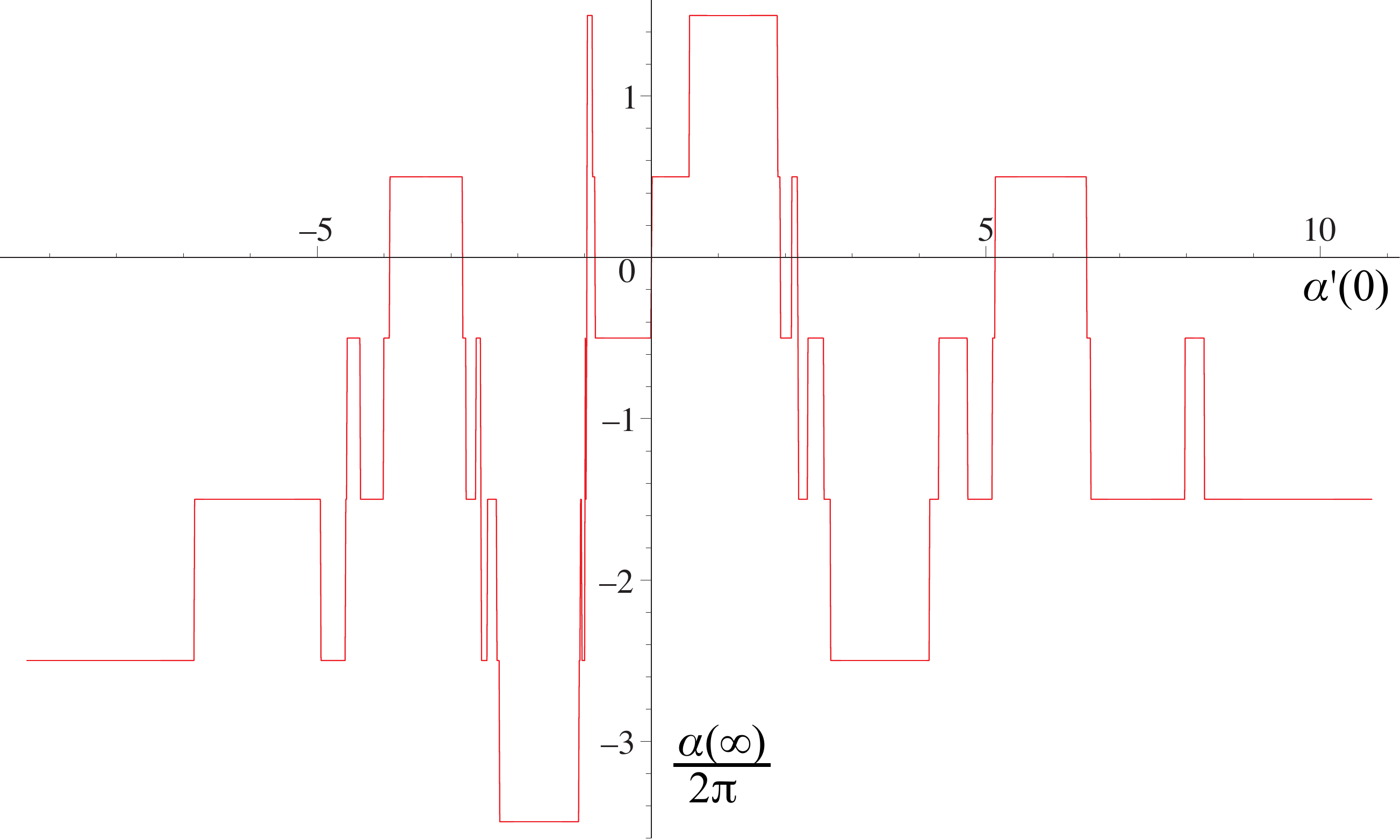}
\caption{Each integer value of this curve implies a Gribov copy of finite norm at a particular $\alpha'(0)$. There is a continuum of solutions with noninteger values, but they have infinite norm. The integer is the winding number of the gauge transformation that takes the reference configuration to the copy. The reference configuration (corresponding to $\alpha'(0)=0$) in this example is $f_1(r)=f_2(r)=(rf_3(r))'/2$ with $f_3(r)=-10r^3e^{-r}$.}\end{figure}

This proliferation of copies also means a proliferation of paths in configurations space that connect copies. If the path is along a gauge orbit then the relative winding number along the path must be noninteger, and as we have seen this yields configurations with infinite norm. If the path is not a gauge orbit then the configurations can retain finite norm and stay within Coulomb gauge, and among such paths are those that correspond to quantum tunnelling events. But paths that are classical real time solutions are another possibility, as we shall now discuss.

\section{Can some transitions be classical?}
We promote the $f_k$'s to functions of $r$ and $t$ and consider their classical evolution. $f_0$ is determined instantaneously on each time slice by Gauss's Law
\begin{equation}
rf_0''+2((1+f_1)^2+f_2^2)f_0/r=2f_2\dot{f}_1-2f_1\dot{f}_2
,\end{equation}
along with the condition $f_0(\infty)=0$. $f_3$ has again been eliminated via (\ref{e1}). Thus a classical solution is described by the pair of functions $(f_1(r,t),f_2(r,t))$. We will keep the $r$ dependence implicit and use the notation $(f_1,f_2)_t$.  We only consider smooth configurations having finite norm and finite action, which requires $(f_1,f_2)_t\rightarrow (0,0)$ for both $r\rightarrow0$ and $r\rightarrow\infty$.

Let $(f_1,f_2)_{t_i}$ specify a configuration on an initial time slice. The first derivatives $(\dot{f}_1,\dot{f}_2)_{t_i}$ then determine the subsequent classical evolution. The total energy is determined by the values of both $(f_1,f_2)_{t_i}$ and $(\dot{f}_1,\dot{f}_2)_{t_i}$ and it is conserved along the classical path. We require that it be finite. For initial data sufficiently smooth and of finite norm, the solutions of the Yang-Mills equations can be extended to all time \cite{a6}.

For a given initial $(f_1,f_2)_{t_i}$ we are interested in the space ${\cal F}[(f_1,f_2)_{t_i}]$ of pairs of $r$ dependent smooth functions that are reachable by classical evolution, i.e. that can appear as $(f_1,f_2)_{t_f}$ for some time $t_f>t_i$. In addition to $t_f-t_i$ this space is parameterized by all choices for the pair of smooth functions $(\dot{f}_1,\dot{f}_2)_{t_i}$. Meanwhile the function pair $(f_1,f_2)_{t_i}$ also determines a set of Gribov copies $(f^{\{j\}}_1,f^{\{j\}}_2)$, $j=1...N$, assumed here to be nonempty. The main question is whether there exists a nonvanishing intersection between these copies and the space of functions ${\cal F}[(f_1,f_2)_{t_i}]$.

If the $j$th copy $(f^{\{j\}}_1,f^{\{j\}}_2)$ is contained in ${\cal F}[(f_1,f_2)_{t_i}]$ for some $(f_1,f_2)_{t_i}$ then there is a classical Minkowski space solution that interpolates between two gauge equivalent spatial configurations, $(f_1,f_2)_{t_i}$ at $t_i$ and $(f^{\{j\}}_1,f^{\{j\}}_2)$ at some later time $t_f$. In other words there is a way to choose $(f_1,f_2)_{t_i}$ and $(\dot{f}_1,\dot{f}_2)_{t_i}$ on the initial slice such that the classical solution encounters a copy. The winding number $n_j$ associated with the $\alpha^{\{j\}}$ gauge transformation yielding this $j$th copy determines the topological charge $q=n_j$ of the interpolating classical solution. We note again that the set of possible $n_j$ for a particular $(f_1,f_2)_{t_i}$ is finite, in contrast to the infinite set of winding number sectors of the non-Coulomb-gauge vacuum.

We can discuss three cases for the nature of the space ${\cal F}[(f_1,f_2)_{t_i}]$ that would determine the likelihood of such transitions. In this discussion we take the initial configuration $(f_1,f_2)_{t_i}$ as fixed.

1) If ${\cal F}[(f_1,f_2)_{t_i}]$ encompasses \textit{all} pairs of smooth functions of finite norm and finite action then there will be a classical solution that connects $(f_1,f_2)_{t_i}$ with any of its copies. The degrees of freedom in the parameterization of ${\cal F}[(f_1,f_2)_{t_i}]$ in terms of $(\dot{f}_1,\dot{f}_2)_{t_i}$ and $t_f-t_i$ indicates that this is in principle possible. There is a related assertion \cite{a8} that the spherically symmetric Yang-Mills system of equations is ergodic, i.e.~that all of phase space is sampled over sufficiently long times. This means that if $t_f-t_i$ is large enough we will eventually run into a copy. This is before making use of the freedom to choose $(\dot{f}_1,\dot{f}_2)_{t_i}$, and so the use of that freedom could presumably produce solutions with a smaller and more relevant $t_f-t_i$.

2) Due to the nonlinearities of the field equations, it is possible that there is some obstruction to ${\cal F}[(f_1,f_2)_{t_i}]$ covering the whole space. Even then it could still be that a random point in the whole space has a nonvanishing probability of being within ${\cal F}[(f_1,f_2)_{t_i}]$. So even if the copies $(f^{\{j\}}_1,f^{\{j\}}_2)$ were nothing more than a random sampling, some may still lie in ${\cal F}[(f_1,f_2)_{t_i}]$.

3) Finally, suppose that a random point in the whole space has a vanishing probability of being within ${\cal F}[(f_1,f_2)_{t_i}]$. Even here one should not conclude that all the copies lie outside, since the set of Gribov copies differs from a random sampling. The quantity $||A^{\{j\}}-A||/||A||$ for example measures how ``close'' the $j$th copy is to the original configuration. As $N$ increases the number of close copies tends to increase, and these close copies can have nonvanishing relative winding numbers.\footnote{For the example of Fig.~(1) there are 11 copies with $||A^{\{j\}}-A||/||A||<1$ and nonvanishing winding numbers.} In addition the closest such copy also tends to become closer as $N$ increases. [In \cite{a2} we were interested in the distribution of the separations between copies having the same winding number, and we found that this distribution was enhanced at small separations.] These properties are related with the increasing density of copies, and in particular the way $N$ increases faster than the range of $\alpha'(0)$ over which copies occur, in the representation of Fig.~(1). We know for sure that ${\cal F}[(f_1,f_2)_{t_i}]$ covers at least some region that is close to the original configuration, and so even in this case some copies may lie within this region.

Which of these three cases applies can also depend on the initial configuration $(f_1,f_2)_{t_i}$, and so the overall result is that we expect at least some Gribov copies with relative winding numbers to be connected by classical solutions.

\section{Conclusions}
Gribov copies are clearly a central feature of the structure of the gauge field configuration space, but the link between their existence and the physical properties of the theory remains to be fully developed. Topological fluctuations have been explored in this paper as one way to make this connection. In section 2 we demonstrated a proliferation of Gribov copies with relative winding number. From section 3 we argued that it is highly unlikely that none of these pairs of copies are connected by classical solutions. Spacetime configurations that do interpolate between Gribov copies with relative winding number are included in the path integral of the quantum theory, and when they are smooth classical solutions of finite norm they are candidates for the dominant localized topological fluctuations. In particular such classical solutions lack a typical tunnelling suppression factor.

Gribov copies are already known to have dramatic effects in general on gluon fluctuations and propagation. These effects can be incorporated explicitly into the path integral, where copies are accounted for by a factor that divides by the number of copies, $1/N(A)$ \cite{a1}. This can be shown to modify the propagator so as to suppress gluon propagation in the strongly coupled infrared \cite{a1,a2}. Another way to describe these effects is that as the number of Gribov copies increase, it becomes more likely that the fields will evolve back to where they were at some earlier time, up to a gauge transformation. This is another way to see that physically nontrivial (gauge inequivalent) propagation effects are suppressed.

But associated with the recurring gauge equivalent configurations there can be changes in winding number. Thus the decrease in the normal gluon fluctuations is correlated with an increase in topological fluctuations. As discussed in \cite{a2}, from the distribution of Gribov copies one can also deduce an enhancement of the infrared spectrum of the Fadeev-Popov operator, which in turn implies an enhancement of the long range Coulomb potential. Both this effect and the suppression of the normal gluon fluctuations are associated with confinement \cite{a1}.\footnote{For another view of the link between confinement and Gribov copies see \cite{a16}.} Thus our discussion here is showing a link between confinement and topological fluctuations.

Our results also call into question the idea that topological fluctuations can be realistically described in terms of instantons alone. Since an instanton is a vacuum to vacuum transition we can characterize an instanton as an incontractible loop where at least one point on the loop is pure-gauge. But the set of such loops is apparently a set of measure zero in the space of all incontractible loops. Thus it is unclear as to why one would expect instantons to dominate the topological fluctuations in strongly interacting gauge theories. This question arises even if incontractible loops include quantum mechanical tunnelling transitions but not real time classical solutions. The question becomes even more pressing if the classical solutions also exist, as we are suggesting here, since they have even more reason to dominate in the path integral.

\section*{Acknowledgments}
I thank R.~Jackiw for bringing reference \cite{a13} to my attention. This work was supported in part by the Natural Science and Engineering Research Council of Canada.

\end{document}